﻿

\documentclass[twocolumn]{aastex62}
\usepackage{amsmath}
\usepackage{graphicx} 
\usepackage{auto-pst-pdf}
\usepackage{rotating}

\shorttitle{Star Formation Properties of BOSS Voids}
\shortauthors{Jian et al.}

\begin{document}

\title{Star Formation Properties of Sloan Digital Sky Survey BOSS Void Galaxies in the Hyper Suprime-Cam Survey}

\correspondingauthor{Hung-Yu Jian}
\email{hyjian@asiaa.sinica.edu.tw}

\author{Hung-Yu Jian}
\affiliation{Institute of Astronomy \& Astrophysics, Academia Sinica, Taipei, 10617, Taiwan}

\author{Lihwai Lin}
\affiliation{Institute of Astronomy \& Astrophysics, Academia Sinica, Taipei, 10617, Taiwan}

\author{Bau-Ching Hsieh}
\affiliation{Institute of Astronomy \& Astrophysics, Academia Sinica, Taipei, 10617, Taiwan}

\author{Kai-Yang Lin}
\affiliation{Institute of Astronomy \& Astrophysics, Academia Sinica, Taipei, 10617, Taiwan}

\author{Keiichi Umetsu}
\affiliation{Institute of Astronomy \& Astrophysics, Academia Sinica, Taipei, 10617, Taiwan}

\author{Carlos Lopez-Coba}
\affiliation{Institute of Astronomy \& Astrophysics, Academia Sinica, Taipei, 10617, Taiwan}


\author{Yusei Koyama}
\affiliation{Subaru Telescope, National Astronomical Observatory of Japan, 650 North A’ohoku Place, Hilo, HI 96720, USA}
\affiliation{Graduate University for Advanced Studies (SOKENDAI),
2-21-1 Osawa, Mitaka, Tokyo 181-8588, Japan}

\author{Chin-Hao Hsu}
\affiliation{Institute of Astronomy \& Astrophysics, Academia Sinica, Taipei, 10617, Taiwan}

\author{Yung-Chau Su}
\affiliation{Institute of Astronomy \& Astrophysics, Academia Sinica, Taipei, 10617, Taiwan}

\author{Yu-Yen Chang}
\affiliation{Department of Physics, National Chung Hsing University, 40227, Taichung, Taiwan}
\affiliation{Institute of Astronomy \& Astrophysics, Academia Sinica, Taipei, 10617, Taiwan}

\author{Tadayuki Kodama}
\affiliation{National Astronomical Observatory of Japan, 2-21-1 Osawa, Mitaka, Tokyo 181-8588, Japan}

\author{Yutaka Komiyama}
\affiliation{National Astronomical Observatory of Japan, 2-21-1 Osawa, Mitaka, Tokyo 181-8588, Japan}
\affiliation{Graduate University for Advanced Studies (SOKENDAI),
2-21-1 Osawa, Mitaka, Tokyo 181-8588, Japan}

\author{Surhud More}
\affiliation{Kavli Institute for the Physics and Mathematics of the Universe (Kavli IPMU, WPI), University of Tokyo, Chiba 277-8582, Japan}

\author{Atsushi J. Nishizawa}
\affiliation{Institute for Advanced Research, Nagoya University, Furocho, Nagoya 464-8602, Japan}

\author{Masamune Oguri}
\affiliation{Research Center for the Early Universe, University of Tokyo, Tokyo 113-0033, Japan}
\affiliation{Department of Physics, University of Tokyo, Tokyo 113-0033, Japan}
\affiliation{Kavli Institute for the Physics and Mathematics of the Universe (Kavli IPMU, WPI), University of Tokyo, Chiba 277-8582, Japan}

\author{Ichi Tanaka}
\affiliation{Subaru Telescope, National Astronomical Observatory of Japan, 650 North A’ohoku Place, Hilo, HI 96720, USA}

\begin{abstract}
We utilize the Hyper Suprime-Cam (HSC) Wide Survey to explore the properties of galaxies located in the voids identified from the Baryon Oscillation Spectroscopic Survey (BOSS) up to $z\sim$0.7. The HSC reaches $\textit{i}\sim$25, allowing us to characterize the void galaxies down to 10$^{9.2}$ solar mass. We find that the revised void galaxy densities, when including faint galaxies in voids defined by bright galaxies, are still underdense compared to the mean density from the entire field. In addition, we classify galaxies into star-forming, quiescent, and green valley populations, and find that void galaxies tend to have slightly higher fractions of star-forming galaxies under the mass and redshift control, although the significance of this result is only moderate (2$\sigma$). However, when we focus on the star-forming population, the distribution of the specific star formation rate (sSFR) of void galaxies shows little difference from that of the control galaxies. Similarly, the median sSFR of star-forming void galaxies is also in good agreement with that of the star-forming control galaxies. Moreover, the effective green valley fraction of void galaxies, defined as the number of green valley galaxies over the number of nonquiescent galaxies, is comparable to that of the control ones, supporting the suggestion that void and control galaxies evolve under similar physical processes and quenching frequencies. Our results thus favor a scenario of the galaxy assembly bias.       
\end{abstract}


\section{Introduction} \label{sec:intro}
Cosmic voids represent vast underdense environments where galaxies barely interact and likely grow in isolation. As a result, void galaxies are less influenced by the transformation processes, such as ram pressure stripping or mergers, that are acting in groups and clusters. The void region is thus an ideal location for studying galaxy secular evolution in the absence of nurture processes and for assessing the environmental impact on the galaxy evolution when compared to galaxies in other environments. 

There has been a broad range of research discussing the environmental influence on galaxy properties in dense surroundings, like clusters and groups. Denser environments result in galaxies being older, redder, and less star-forming than those in less dense conditions \citep{dre80,bal04,coo07,ger07,pen10,muz12,wet12,lin14,jian17,jian18,jian20}. Some works have shown that while the fraction of quiescent (or star-forming) galaxies depends on their hosting environment, the star-forming main sequence exhibits weak or no environmental correlation \citep{bal04,vul10,koy13,lin14}, supporting a fast environmental quenching scenario. Contrarily, studies have indicated that in addition to the apparent environmental dependence of the quiescent (or star-forming) fraction, the star-forming main sequence in galaxy clusters shows a global reduction of the specific star formation rate
(sSFR) as opposed to the field, implying a slow environmental quenching effect acting in extreme environments \citep{vul10,hai13,alb14,lin14,jian17,jian18,jian20}.  

Investigations in underdense environments have also led to diverse conclusions on the environmental dependence of galaxy properties. Some studies show that the properties of void galaxies, e.g., color, morphology, sSFR, or gas mass, appear to show little difference from those of typical galaxies. For example, \cite{pat06} indicate that when probing blue and red galaxies separately, the color distributions, morphology distributions, and correlations of sSFR with color seem to be nearly identical between the void and overall galaxies. \cite{kre12} make use of 60 void galaxies and find no increase of the $H_I$ mass-to-light ratio in low-density regions, and, in general, observe very few void galaxies with elevated $H_I$ mass-to-light ratios.

On the other hand, other works have shown that void galaxies are bluer, less evolved, and have more star-forming galaxies than the control sample. \cite{hoy12} show that with the same magnitude distribution, void galaxies statistically display bluer color than those in overdense regions. Under the morphology control, late-type (early-type) void galaxies reveal bluer color than late-type (early-type) control galaxies. \cite{ric14} find that passive void galaxies are strongly suppressed compared to the control sample. Even when the mass is under control, galaxies in voids statistically exhibit different sSFRs from those in the general population. \cite{bru20} find a deficit of luminous and normal red galaxies in voids, resulting in a mean (g – r) magnitude color difference between the wall and void galaxies, but no significant color difference when the blue and red samples are considered separately. Recently, \cite{flo21} study void galaxies by selecting the 10$\%$ of galaxies with the lowest densities, where the density is a function of the distance to the third-nearest neighbor. They find that at fixed stellar mass, void galaxies are bluer, more gas-rich, and more star-forming than control galaxies. They thus suggest that these trends are possible signatures of galaxy assembly bias.        

This work combines the void catalog from the Baryon Oscillation Spectroscopic Survey \citep[BOSS;][]{mao17} with the wide-field galaxy dataset from the Hyper Suprime-Cam Survey \citep[HSC;][]{aih18,aih19} to explore void galaxy properties. Because the Sloan Digital Sky Survey (SDSS) DR12 BOSS voids were identified with galaxies brighter than $i \leq 19.9$, the properties of faint void galaxies have not been not well explored. In this work, we utilize galaxy stacking of the BOSS voids combined with the background/foreground subtraction technique in the fields that are overlapped with the HSC survey to obtain statistical results for properties of faint void galaxies with $\textit{i} \leq 25$. Compared to previous works, we thus can probe void galaxies with less stellar mass and higher redshift up to 0.7 under the mass and redshift control. In addition, we can also answer whether a ``revised void'' by the inclusion of faint galaxies in a void previously defined by bright galaxies, remains an underdense region. Our results may thus shed light on the role of the environmental effect in galaxy evolution.

We outline the remaining sections of this paper as follows. In Section 2, we briefly describe the data set and our sample selection, and we illustrate the analysis methods in Section 3. Section 4 presents the main results, including discussions of the density map of star formation rate (SFR)\textendash $M_*$, the distribution of sSFR, and the redshift and mass dependence of the quiescent star-forming and green valley galaxy fractions in voids and the control sample. Finally, conclusions are given in Section 5. Throughout this paper, we adopt a flat $\Lambda$ cold dark matter cosmology with the following parameters: $H_0$ = 100h km s$^{-1}$ Mpc$^{-1}$ , $\Omega_{m}$ = 0.3, and $\Omega_{\Lambda}$= 0.7. We adopt the Hubble constant $h$ = 0.7 in the calculation of rest-frame magnitudes. All magnitudes are in the AB system.

\begin{figure*}
\includegraphics[scale=1.4]{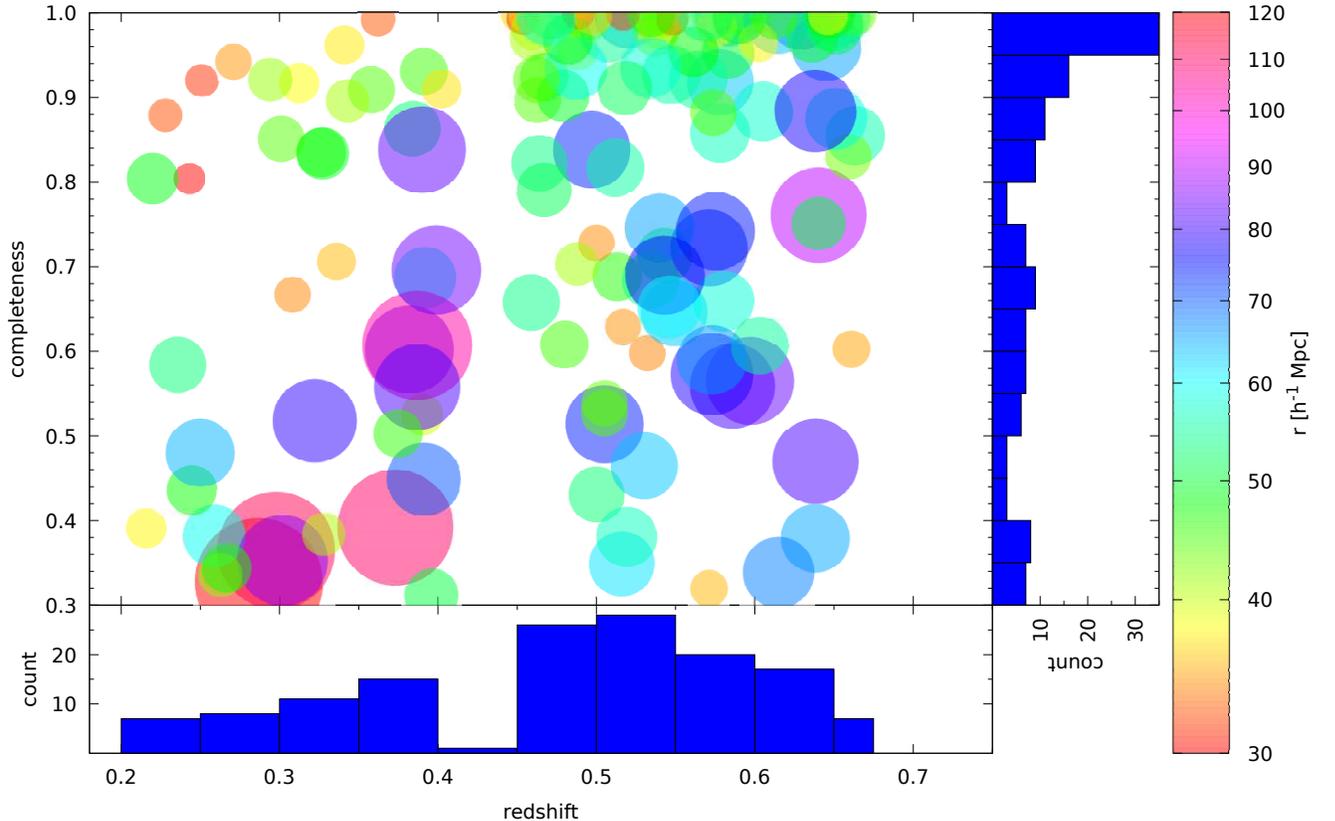}
\caption{The void sample is shown as functions of redshift, area completeness, and size, where the area completeness is defined as the ratio of the overlapping area between the HSC field and a BOSS void to the entire void area. The histograms show the number counts of redshift and completeness, respectively. The sizes of the circles are color-coded and also proportional to the void radius. We select voids with radius $\geq$ 30 $h^{-1}$ Mpc and their area completeness $\geq$ 0.3 in redshift range between 0.2 and 0.7. The final sample used in this study consists of 69 and 70 voids in 0.2 $< z <$ 0.5 and 0.5 $< z <$ 0.7, respectively.}   
\label{f1}
\end{figure*}

\section{Data and Sample Selection}

This work is based on the S19A internal data release, including imaging data observed from 2014 March to 2019 May, covering $\sim$1200 deg$^2$, from the HSC Subaru Strategic Program \citep[][]{aih18,aih19}. The S19A data release is processed with the updated pipeline hscPipe \citep[][version 7]{bos18}, which is a version of the pipeline for the Vera C. Rubin Observatory \citep{jur17,ive19} with HSC-specific features included. Calibration for the photometry and astrometry is done against the Pan-STARRS1 $3\pi$ catalog \citep{sch12,ton12,mag13,cha16}, which is further calibrated against Gaia DR1. The quality assurance tests of the S19A data release show that the astrometry is as good as 10-30 milliarcsec against GAIA, and photometry is good to $\sim$ 0.01-0.02 mag. 

Our sample is selected from the wide-layer data in the S19A release, which reaches $\textit{i} \sim$ 26.2 \citep{aih21}. We further restrict galaxies to be full color, i.e. all $\textit{grizy}$ bands are detected, and have a flux limit cut at $\textit{i}$ = 25. In total, we have roughly 140 million galaxies in the field of $\sim$900 deg$^2$ in our galaxy dataset.   


The S19A internal data release also includes photometric redshift catalogs, which provide photometric redshift, stellar mass, and other derived physical quantities. In this work, the estimations of the photometric redshifts and stellar masses are derived using the direct empirical photometric method \citep[DEmP;][]{hsi14}. DEmP adopts ``regional polynomial fitting'' to select a local subset dynamically to derive the relation represented by a first-order polynomial function between redshift and photometry for a given input galaxy. In this manner, a random variation in the color\textendash magnitude space can be represented by multiple line segments without the need to choose a complicated fit function that satisfies all data. DEmP thus reduces the selection effects of choosing the proper form for the fit function and the bias in the best-fit coefficients. The detailed description is in \cite{hsi14}. Based on the S19A photo$z$ release note, an inaccuracy of $\sim$0.055 is quoted for DEmP photo$z$, and DEmP stellar mass, trained against COSMOS masses, shows a mean ($\Delta$log$M_*$ = log$M^{\textrm{hsc}}_*$ - log$M^{\textrm{cosmos}}_*$) of -0.02 dex and a scatter of 0.2 dex.



Our void sample is adopted from the cosmic void catalog based on the final galaxy catalog of SDSS-III from BOSS \citep{mao17}. They utilized the ZOBOV void-finding algorithm \citep{ney08} to search for voids, and identified a total of 10,643 voids. Among those, 1228 voids were left after quality cuts. They found that the effective radii of these 1228 voids range from 20 to 100 $h^{-1}$ Mpc, with the density in the central region being 30$\%$ of the mean density from the entire field. The detailed information is described in \cite{mao17}.

Because the sky coverage of HSC is much smaller than that of BOSS, not every BOSS void is under the HSC footprint. Besides, the overlapping region between some BOSS voids and the HSC field is small. Therefore, we restrict our void sample to consist of voids with area completeness larger than 30$\%$. We further select voids with radii larger than 30 $h^{-1}$ Mpc and split the voids into two redshift bins, namely, into redshift ranges of 0.2 $< z <$ 0.5 and 0.5 $< z <$ 0.7. Finally, we obtain 69 and 70 voids at low and high redshift, respectively. We plot the void sample as functions of redshift, area completeness, and color-coded size in Figure~\ref{f1}. A redshift gap at $z\sim$0.4 in the void sample can be seen in Figure~\ref{f1}, which is also found in the original BOSS void sample, and is not due to our sample selection.      
 
\begin{figure*}
\includegraphics[angle=90,scale=0.6]{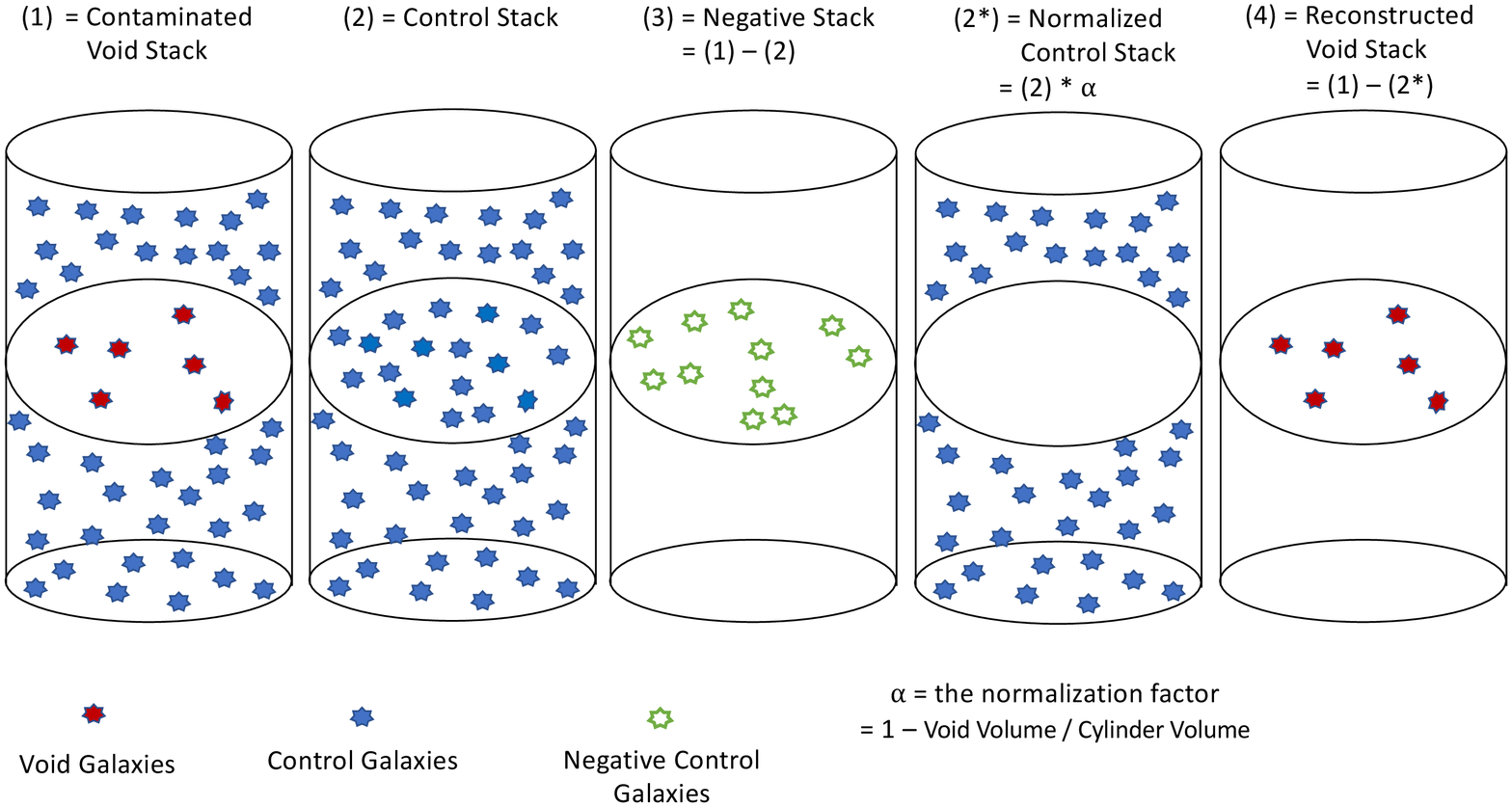}
\caption{A schematic to demonstrate the different stacks in our analysis. (1) Contaminated void stack: void plus foreground and/or background galaxies. (2) Control stack: all galaxies in the same redshift range of the corresponding void and normalized to the size of the void stack. (3) Negative stack: direct subtraction of the control stack from the contaminated void stack, i.e. (1)$-$(2). (2$^*$) Normalized control stack: control stack times a normalization factor defined as 1 $-$ void volume/cylinder volume. (4) Reconstructed void stack: subtraction of the normalized stack from the contaminated void stack.}
\label{f2}
\end{figure*}

\section{Methods}\label{method}

Following the approach described in \citet{jian18,jian20}, we perform galaxy stacking combined with the background/foreground subtraction technique to obtain the statistical properties of void galaxies. Figure~\ref{f2} presents a
schematic diagram illustrating the different stacks and the operations we conduct in our analysis to acquire the reconstructed void sample at the end. For each void, galaxies are projected onto a 2D plane around the void center. We restrict the projected galaxies above a flux limit and within a redshift slice. The width of the redshift slice is limited between 1 $\sigma$ above and below the void redshift, where $\sigma$ is the photo$z$ dispersion of the galaxy sample defined in \cite{tan18} as $\sigma$ = 1.48 $\times$ MAD($\Delta z$), where $\Delta z$ = $z_{\textrm{phot}}$–$z_{\textrm{spec}}$ and MAD is the median absolute deviation, which is approximately 0.055 in this work. 

These projected galaxies within one void radius constitute a one-void contaminated stack, which constitutes both void members and foreground/background galaxies. We accumulate all one-void contaminated stacks with the radius normalization to establish the contaminated void stack, i.e. (1) in Figure~\ref{f2}. Similarly, we take all of the projected galaxies from the entire survey within the same redshift slice normalized to the corresponding projected void area to form a one-void control stack. We note that the final control sample is a mixture of galaxies from all environments, including clusters, groups, the field, and voids. Therefore, the signatures of the reconstructed void stack are likely to be partially smeared out when compared to our control sample, leading to a weaker signal and/or confidence level. Combining all of the one-void control stacks, we have the control stack, i.e. (2) in Figure~\ref{f2}. When we perform the direct background subtraction, i.e., subtract the control sample (2) from the contaminated void stack (1), we expect to obtain a negative sample (3) due to the lower density environment in (1) than in (2). Although the negative sample (3), directly presents a negative signal, the interpretation for (3) is difficult. Instead, we construct a one-void normalized stack by multiplying a one-void control sample (2) and a normalization factor $\alpha$, defined as 1$-$void volume/cylinder volume, to build a sample consisting of galaxies inside the cylinder but not in the void size region. The stacking of all f the one-void normalized stacks is the normalized stack (2$^*$). Finally, we subtract the normalized stack (2$^*$) from (1) to obtain the reconstructed void sample (4). In practice, for a given parameter space (e.g., SFR vs. stellar mass, or SFR vs. $M_{*}$), we compute the number counts in each cell on the plane for the contaminated void and normalized control samples separately. We then recover the statistical properties of void galaxies by subtracting the number count in a cell of the control sample from the number count in the corresponding cell of the contaminated void sample.

We follow the procedures described in \cite{jian18,jian20} to compute the SFRs for the projected galaxies. In short, we substitute individual photo$z$s for the same void redshift during stacking, and compute the rest-frame $B$ magnitude $M_{B}$ and $(U-B)_0$ color for each galaxy by using the K-correction based on empirical templates from \cite{kin96}. We then employ the best-fit results listed in Table 3 of \cite{mos12} of an empirical polynomial formula using the rest-frame $B$ magnitude, $(U-B)_0$ color and second-order $(U-B)_0$ color as fitting parameters to estimate the SFR. The SFR estimation is calibrated against the spectral energy distribution\textendash fitted SFRs from UV/optical bands in the All-Wavelength Extended Groth Strip International Survey (AEGIS) for both quiescent and star-forming galaxies in the redshift range of 0.7 $< z <$ 1.4. This calibration yields an uncertainty of approximately 0.19 dex for star-forming and 0.47 dex for quiescent galaxies, with a mean residual offset of $-$0.02 dex. Besides, \cite{mos12} made a comparison between their multicolor SFR calibration to sL[O$_{\textrm{II}}$]-base SFR calibrations from local samples, and concluded that they are in good agreement. 

To estimate the stellar mass completeness limit, we adopt the method from \cite{ilb10}. Using the HSC dataset, we compute the fraction of galaxies as a function of stellar mass at an $\textit{i}$=25 cut in relation to the full sample ($\textit{i}$=26). Then we define the low stellar mass as the mass at which  30\% of the galaxies are fainter than $\textit{i}$=25. Following this approach, we find that in our data set the mass limits are log$_{10}$($M_*$ /$M_\odot$) = 8.5 (9.2) and 9.2 (9.8) for star-forming (quiescent) galaxies in redshift ranges of 0.2\textendash 0.5 and 0.5\textendash 0.7, respectively.

\begin{deluxetable*}{ccccccccc}
\tablenum{1}
\tablecaption{Best-fit Parameters for the Star-forming Main Sequence, Red Sequence, and Green Valley\label{tab1}}
\tablehead{
\colhead{Redshift} & \multicolumn{2}{c}{Star-forming Main Sequence} & \colhead{} & \multicolumn{2}{c}{Red Sequence}  &  \colhead{} & \multicolumn{2}{c}{Green Valley} \\ 
\cline{2-3}\cline{5-6}\cline{8-9}
\colhead{} & \colhead{$\alpha$\tablenotemark{a}} & \colhead{$\beta$\tablenotemark{a}} &  \colhead{} & \colhead{$\alpha$\tablenotemark{a}} & \colhead{$\beta$\tablenotemark{a}} &  \colhead{} & \colhead{$\alpha$\tablenotemark{a}} & \colhead{$\beta$\tablenotemark{a}}    
}

\startdata
 0.2 $< z <$ 0.5   & 0.83 $\pm$ 0.01 & -7.68 $\pm$ 0.12  & & 0.75 $\pm$ 0.01 & -8.31 $\pm$ 0.13  & & 0.79 $\pm$ 0.02 & -7.99 $\pm$ 0.18  \\
 0.5 $< z <$ 0.7   & 0.82 $\pm$ 0.01 & -7.26 $\pm$ 0.11  & & 0.60 $\pm$ 0.01 & -6.43 $\pm$ 0.06  & & 0.71 $\pm$ 0.02 & -6.90 $\pm$ 0.13 \\
\enddata
\tablenotetext{a}{$\alpha$ and $\beta$ are the fitting slope and amplitude for the fitting formula, $log_{10}(\rm{SFR}/M_{\odot}  ~ yr^{-1}) = \alpha ~ log_{10}(M_*/M_{\odot}) + \beta$, respectively.}
\end{deluxetable*}

\begin{figure*}
\centering

\includegraphics[scale=1.6]{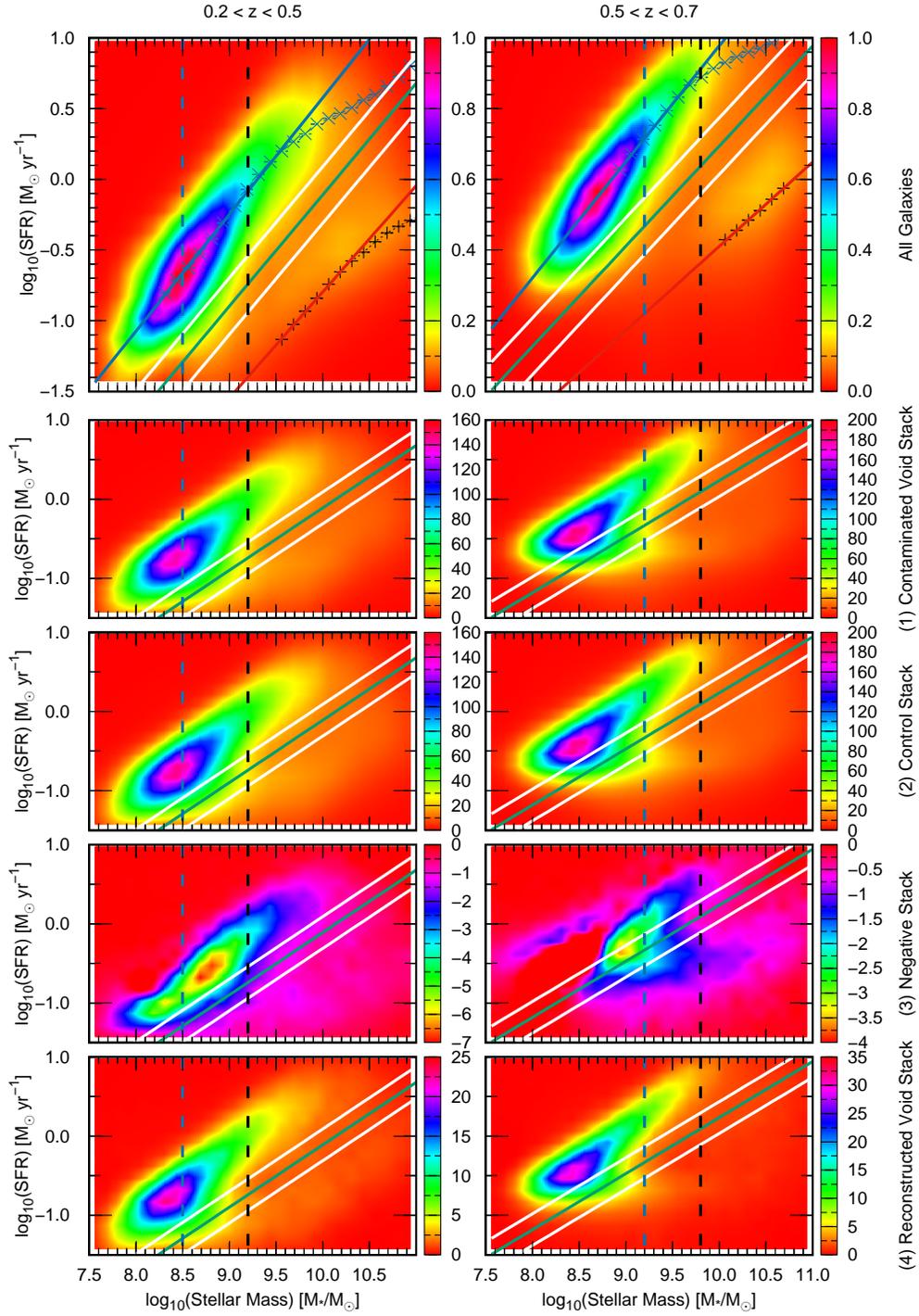}
\caption{Color-coded density plots from the different sample stacks. Top: the stack of all galaxies with good photo$z$, i.e., the deviation of photo$z$ $\leq$ 0.3. The SFR estimation uses the individual photo-$z$ of every galaxy, and the cell density is normalized to the maximum density. Bottom: (1) the contaminated void stack; (2) the control stack; (3) the negative stack; and (4) the reconstructed void stack, corresponding to (1) to (4) in Figure~\ref{f2}, respectively. It should be noted that from (1) to (4), the galaxy SFR is estimated using the BOSS void redshift, and the density is scaled down by dividing by 10$^{3}$.}

\label{f3}
\end{figure*}

\begin{figure}

\includegraphics[scale=0.65]{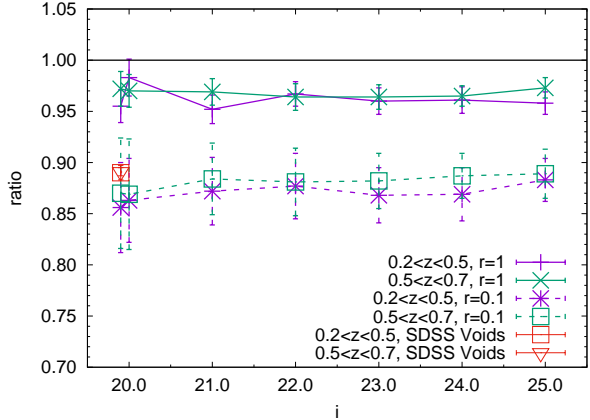}
\caption{A density ratio plot of the contaminated void to control stacks, i.e. (1)/(2) in Figure~\ref{f3}, as a function of $i$-band magnitude cut. The purple and green lines represent the results at high and low redshift, respectively, while the solid line denotes the size of one void radius, and the dashed line marks that of one-tenth void radius. The red open square (high-$z$) and triangle (low-$z$) give the derived BOSS values using the core density. The density ratio appears to be independent of the magnitude cut and redshift, but correlates with the radial position within the void. The independence of the magnitude cut implies that including faint galaxies for a void defined by bright galaxies will not change its underdense status.}
\label{f4}
\end{figure}

\begin{deluxetable*}{ccccccccccccc}
\tablenum{2}
\tablecaption{Properties as in Figure~\ref{f6}} \label{tab2}
\tablehead{
\colhead{Redshift} & \colhead{} & \colhead{Environment} & \colhead{} & \colhead{Mass Range} & \colhead {} &  \colhead{Quiescent} & \colhead{} & \colhead{Star-forming} & \colhead{} &  \colhead{Green} & \colhead{} & \colhead{Total Galaxy Number\tablenotemark{a}}\\
\cline{1-1} \cline{3-3} \cline{5-5} \cline{7-7} \cline{9-9} \cline{11-11} \cline{13-13}
\colhead{} & \colhead{} & \colhead{} & \colhead{} & \colhead{$log_{10}(M_*/M_{\odot})$} & \colhead{} & \colhead{$f_q$\tablenotemark{b}} & \colhead{} & \colhead{$f_s$\tablenotemark{b}} & \colhead{} & \colhead{$f_g$\tablenotemark{b}} & \colhead{} & \colhead{\#} 
}

\startdata
  & &   & & 9.5-9.75 & &  0.275$\pm$0.035 & & 0.586$\pm$0.040 & &  0.139$\pm$0.012  & &  6.495$\times 10^{4}$    \\ 
  & &   & & 9.75-10.0 & & 0.350$\pm$0.032 & & 0.493$\pm$0.037 & &  0.156$\pm$0.013  & &  5.404$\times 10^{4}$    \\  
  & &   & & 10.0-10.25 & & 0.404$\pm$0.040 & & 0.408$\pm$0.043 & &  0.188$\pm$0.010  & &  3.897$\times 10^{4}$    \\
0.2 $< z <$ 0.5  & & Voids   & & 10.25-10.5 & & 0.448$\pm$0.057 & & 0.335$\pm$0.047 & &  0.217$\pm$0.014  & &  3.079$\times 10^{4}$    \\
  & &   & & 10.5-10.75 & & 0.597$\pm$0.063 & & 0.185$\pm$0.036 & &  0.218$\pm$0.028  & &  2.253$\times 10^{4}$    \\
  & &   & & 10.75-11.0 & & 0.737$\pm$0.058 & & 0.083$\pm$0.023 & &  0.180$\pm$0.037  & &  1.400$\times 10^{4}$    \\  
  \cline{1-13}
  & &   & & 9.5-9.75 & &  0.308$\pm$0.001 & & 0.545$\pm$0.002 & &  0.147$\pm$0.002  & &  4.834$\times 10^{5}$    \\ 
  & &   & & 9.75-10.0 & & 0.382$\pm$0.003 & & 0.464$\pm$0.002 & &  0.154$\pm$0.002  & &  3.980$\times 10^{5}$    \\  
  & &   & & 10.0-10.25 & & 0.471$\pm$0.005 & & 0.349$\pm$0.003 & &  0.180$\pm$0.003  & &  2.990$\times 10^{5}$    \\
0.2 $< z <$ 0.5  & & Control Sample & & 10.25-10.5 & & 0.550$\pm$0.006 & & 0.246$\pm$0.003 & &  0.203$\pm$0.003  & &  2.316$\times 10^{5}$    \\
  & &   & & 10.5-10.75 & & 0.687$\pm$0.004 & & 0.130$\pm$0.004 & &  0.183$\pm$0.002  & &  1.679$\times 10^{5}$    \\
  & &   & & 10.75-11.0 & & 0.830$\pm$0.003 & & 0.049$\pm$0.003 & &  0.121$\pm$0.002  & &  1.067$\times 10^{5}$    \\    
 \cline{1-13}  
  & &   & & 10.0-10.25 & & 0.412$\pm$0.037 & & 0.396$\pm$0.034 & &  0.192$\pm$0.009  & &  4.232$\times 10^{5}$    \\
  & &  & & 10.25-10.5 & & 0.512$\pm$0.041 & & 0.270$\pm$0.030 & &  0.218$\pm$0.013  & &  3.483$\times 10^{5}$    \\
0.5 $< z <$ 0.7  & & Voids  & & 10.5-10.75 & & 0.183$\pm$0.044 & & 0.185$\pm$0.025 & &  0.232$\pm$0.021  & &  2.664$\times 10^{5}$    \\
  & &   & & 10.75-11.0 & & 0.672$\pm$0.050 & & 0.110$\pm$0.023 & &  0.217$\pm$0.030  & &  1.686$\times 10^{5}$    \\    
  \cline{1-13}
  & &   & & 10.0-10.25 & & 0.501$\pm$0.008 & & 0.323$\pm$0.035 & &  0.006$\pm$0.006  & &  3.324$\times 10^{5}$    \\
  & &  & & 10.25-10.5 & & 0.596$\pm$0.009 & & 0.214$\pm$0.040 & &  0.006$\pm$0.015  & &  2.699$\times 10^{5}$    \\
0.5 $< z <$ 0.7  & & Control Sample  & & 10.5-10.75 & & 0.682$\pm$0.010 & & 0.137$\pm$0.005 & &  0.181$\pm$0.022  & &  2.099$\times 10^{5}$    \\
  & &   & & 10.75-11.0 & & 0.778$\pm$0.010 & & 0.082$\pm$0.072 & &  0.150$\pm$0.004  & &  1.338$\times 10^{5}$    \\  
 \enddata
\tablenotetext{a}{``Total galaxy number'' refers to the number of galaxies after background subtraction for void galaxies and the void size-normalized number of galaxies for control galaxies.}
\tablenotetext{b}{$f_q$: quiescent fraction; $f_s$: star-forming fraction; and $f_g$: green valley galaxy fraction.}
\end{deluxetable*}

\section{Results and Discussion}
\subsection{Density Plots of SFR\textendash \texorpdfstring{$M_{*}$}{*}}\label{denplot}
Before we start our analysis, clear definitions of our galaxy classifications, i.e., the types of star-forming, green valley, and quiescent galaxies, in this work are necessary. We follow the method described in \cite{jian20}. In short, we stack all galaxies with ``photoz\_risk\_best'' $\leq$ 0.3, defined as the risk of photoz\_best being outside of the range z\_true $\pm$ 0.15(1+z\_true) ranging from 0 (safe) to 1(risky), on the SFR\textendash $M_*$ plane, and normalize the cell number density by the maximum density in all cells, referring to the top two panels in Figure~\ref{f3}. Please note that galaxy SFRs in these two panels are estimated using their photo$z$s, unlike void galaxy SFRs computed using the same corresponding BOSS void redshift. 

Following the procedures in \cite{jian20}, in the beginning, we separate the star-forming and quiescent galaxies, using a constant log$_{10}$(sSFR) = -10.1, to obtain an initial star-forming main sequence and red sequence, respectively. We note that our result is not sensitive to the initial sSFR value \citep{jian20}. We then define the green valley line as the middle point of these two sequences, and use the green valley line as the new separation line to classify the star-forming and quiescent galaxies. We iterate the process until the green valley line converges and remains unchanged. The green valley is then to be the region, represented by the two solid white lines, between 0.2 dex above and below the green valley line, which is represented by the solid blue line in Figure~\ref{f3}. We verify that the upper (lower) boundary line of the green valley is close to the boundary where star-forming (quiescent) galaxies are roughly 2$\sigma$ away from the star-forming (quiescent) sequence. In this work, we then split our void galaxies into three populations, i.e., star-forming, quiescent, and green valley galaxies, based on their location relative to the green valley. That is, galaxies above (below) the green valley are star-forming (quiescent) galaxies.

Additionally, in Figure~\ref{f3}, we mark the star-forming main sequence with the solid blue line, and the red sequence with the solid red line, respectively. The best-fit coefficients for the star-forming main sequence, the green valley line, and the red sequence in Table~\ref{tab1}. Our best-fit results of the star forming main and quiescent sequences are consistent with those found in \cite{jian18,jian20}. The vertical blue and black dashed lines denote the mass completeness limits for star-forming and quiescent galaxies, respectively.  

The remaining bottom panels (1), (2), (3), and (4) in Figure~\ref{f3} are the contaminated void, control, negative, and reconstructed void stacks, corresponding to (1) to (4) in Figure~\ref{f2}, respectively. Please note that the color-coded density in these four plots is scaled down by dividing 10$^{3}$ for clarity. In panels (3) of Figure~\ref{f3}, we can see the negative density after the background subtraction of (2) from (1) as a result of an underdense void environment respective to the overall mean density. However, direct interpretation for the negative signals is challenging. Instead, as we explain earlier in Figure~\ref{f2}, we can normalize the control stack by the volume normalization factor $\alpha$ to obtain a normalized control stack similar to (2$^*$) of Figure~\ref{f2}. In this way, we recover the underlying member galaxies of voids, i.e. (4). 

\subsection{Density Ratio of Void Stack to Control Stack}

As a simple test to examine whether a stack contains voids, we compute the density ratio of the contaminated void stack (1) to the control stack (2). If the density ratio is less than 1, it indicates that the density of the (1) stack is less than that of the mean-field. That is, the (1) stack contains voids. Figure~\ref{f4} shows the ratio as a function of $i$ band magnitude cut at low (purple) and high (green) redshift, and inside 0.1 (dotted) and 1 (solid) void radius, respectively. In this work, the error bars are estimated using a bootstrap resampling for 200 tries on the void sample, and are the standard deviations of the mean. 

We find that the density ratio is independent of $i$-band magnitude and redshift, but shows a strong correlation with the void-centric radius. The density ratio shows no dependence on the magnitude cut, indicating that a revised void including faint galaxies in the same area of a void density defined by bright galaxies remains an underdense region with the similar density ratio. The ratios are all less than 1, implying that the (1) stack includes voids. The dependence of the ratio on the void-centric radius is understandable, since the density at the void center is likely the least. Thus, the ratio for the stack around the void center also likely reveals the minimum. Additionally, we create a fake void catalog by replacing the original 1228 BOSS void positions with random ones in the HSC field to compute the ratio. The result shows that the ratio is 0.99 $\pm$ 0.006, indicating the density of a random stack is close to the mean density and ensuring that the (1) stack is different from a random catalog and contains void regions.

\begin{figure*}
\includegraphics[scale=1.35]{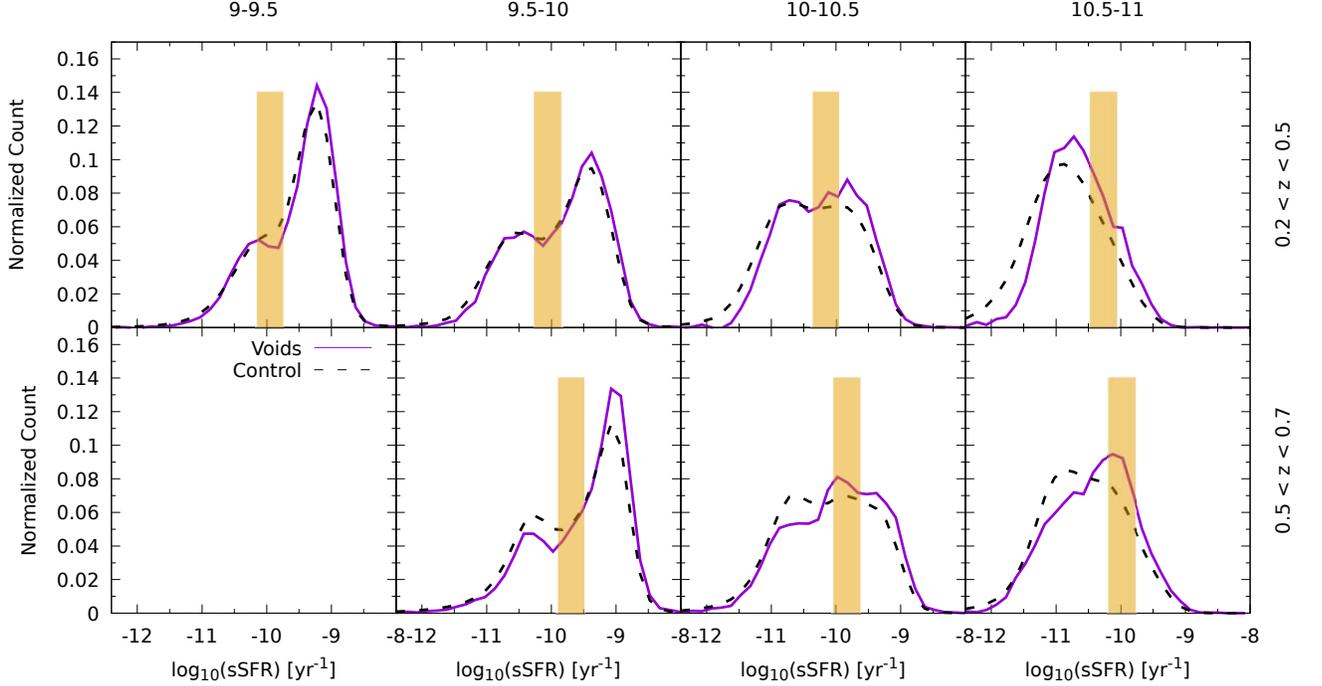}
\caption{The sSFR distributions of void galaxies is displayed as functions of stellar mass and redshift. The solid purple and dashed black lines mark the sSFR distributions of the void and control galaxies, respectively. The golden bars represent the green valley. The two distributions are area-normalized and show similarity, especially at low stellar mass bins.}
\label{f5}
\end{figure*}

\begin{figure*}
\includegraphics[scale=1.3]{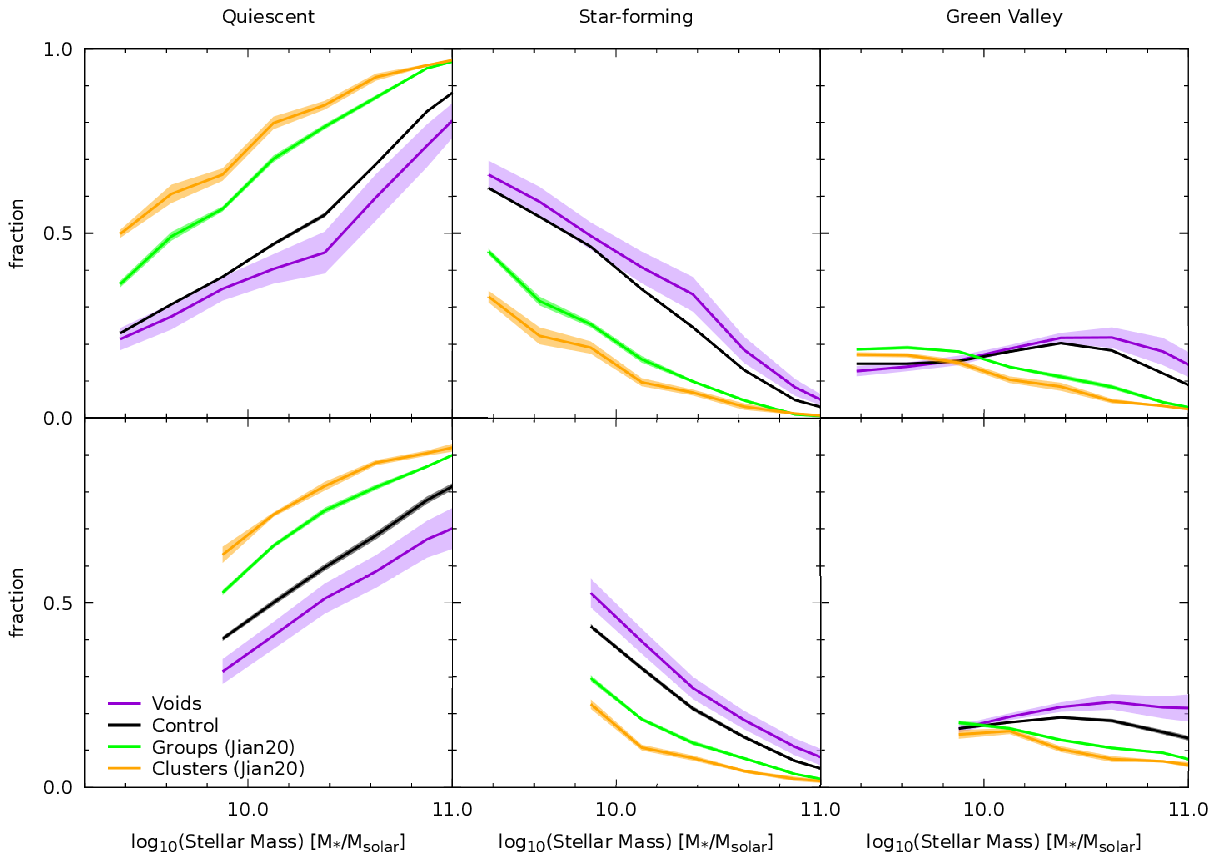}
\caption{The fractions of quiescent ($f_{q}$), star-forming ($f_{s}$), and green valley galaxies ($f_{g}$) from voids (purple), the control sample (black), groups (green), and clusters (gold). The results for groups and clusters are from \cite{jian20}. Error bars are 1$\sigma$ confidence intervals made from a bootstrap resampling. Among the four different environments, the $f_{q}$ of clusters is highest, while voids show the lowest $f_{q}$. In between are the $f_{q}$s of groups and the control sample. In contrast, the $f_{s}$ and $f_{g}$ reveal an opposite trend.}
\label{f6}
\end{figure*}

The BOSS void catalog also provides the density contrast $\delta$, defined as the density contrast of the minimum density cell compared to the mean density at that redshift. We find that the density ratio is connected to $\delta$ as follows.
The ratio of the (1) stack to the (2) stack is equal to
\begin{equation}\label{eq1}
\begin{split}
\frac{\rho_{v'}}{\rho_m}= \frac{N_{v'}/V_{v'}}{N_m/V_m} =\frac{N_{v'}}{N_m} 
=\frac{N_{m}*\alpha + N_v}{N_m}=\alpha+\frac{N_v}{N_m}, 
\end{split}
\end{equation}
where $N_{v'}$, $N_{m}$, and $N_v$ are the galaxy numbers in the (1) stack, (2) stack, and voids, respectively;  $V_{v'}$ = $V_{m}$ is the cylinder volume; and $\alpha$ = 1 - $V_v/V_m$ is the normalization factor defined in (2$^*$) of Figure~\ref{f2}.
In addition, the density contrast is defined as 

\begin{equation}\label{eq2}
\delta = \frac{\rho_v}{\rho_m} - 1 = \frac{N_v/V_c}{N_m/V_m} - 1=\frac{N_{v}}{N_m}\frac{V_m}{V_{v}}-1.
\end{equation}
From Eqn.~\ref{eq2}, we obtain
\begin{equation}\label{eq3}
\frac{N_v}{N_{m}} = (1+\delta)\frac{V_v}{V_m} = (1+\delta)(1-\alpha).
\end{equation}
Substituting the term of $N_v$/$N_m$ in Eqn.~\ref{eq1} with that in Eqn.~\ref{f3}, the ratio becomes
\begin{equation}\label{eq4}
\frac{\rho_{v'}}{\rho_m}= \alpha+\frac{N_v}{N_m}=  \alpha + (1+\delta)(1-\alpha) = 1 + \delta(1-\alpha)
\end{equation}
Using Equation~\ref{eq4}, we convert the $\delta$ of voids from the BOSS void catalog to the density ratios, and obtain the mean and the bootstrap errors for voids at low (the red open square) and high redshift (the red open inverse triangle), respectively, in Figure~\ref{f4}. Our results using one-tenth void radius, referring to the central void stack ratio, appear to agree well with the BOSS results, where the density ratio of a BOSS void is derived from the minimum density of the Voronoi cells in that void. Additionally, two BOSS results also reveal redshift independence, also consistent with our finding.


\begin{deluxetable*}{ccccccccccccccccc}
\tablenum{3}
\tablecaption{Results of the Two-side K\textendash S Test as in Figure~\ref{f5}, Figure~\ref{f7}, and Figure~\ref{f8}} \label{tab3}
\tablehead{
\colhead{} & \colhead{} & \colhead{} & \colhead{} & \colhead{} & \colhead{} & \colhead{Figure~\ref{f5}} & \colhead{} & \colhead{Figure~\ref{f5}} & \colhead{} & \colhead{Figure~\ref{f7}} & \colhead{} & \colhead{Figure~\ref{f7}} & \colhead{} & \colhead{Figure~\ref{f8}} & \colhead{} & \colhead{Figure~\ref{f8}} \\
\cline{7-9} \cline{11-13} \cline{15-17} 
\colhead{Redshift} & \colhead{} & \colhead{Environment} & \colhead{} & \colhead{Mass Range} & \colhead {} &  \colhead{Statistic\tablenotemark{a}} & \colhead{} & \colhead{$p$-value\tablenotemark{b}}  & \colhead {} &  \colhead{Statistic\tablenotemark{a}} & \colhead{} & \colhead{$p$-value\tablenotemark{b}}  & \colhead{} &  \colhead{Statistic\tablenotemark{a}} & \colhead{} & \colhead{$p$-value\tablenotemark{b}}\\
\cline{1-1} \cline{3-3} \cline{5-5} \cline{7-7} \cline{9-9} \cline{11-11} \cline{13-13} \cline{15-15} \cline{17-17}
\colhead{} & \colhead{} & \colhead{} & \colhead{} & \colhead{$log_{10}(M_*/M_{\odot})$} & \colhead{} & \colhead{} & \colhead{} & \colhead{} & \colhead{} & \colhead{} & \colhead{} & \colhead{} & \colhead{} & \colhead{} & \colhead{} & \colhead{} 
}

\startdata
                           & &                & &  9.0$-$9.5   & &  0.450 & &  $\ll$ 0.001 & &  0.150 & &   0.723 & &  0.467 & &  0.051 \\
                           & & Voids          & &  9.5$-$10.0  & &  0.450 & &  $\ll$ 0.001 & &  0.200 & &   0.361  & &  0.500 & &   0.023\\
  0.2 $< z <$ 0.5 & & vs.                     & & 10.0$-$10.5  & &  0.450 & &  $\ll$ 0.001 & &  0.200 & &  0.361 & &  0.412 & &  0.081 \\
                           & & Control Sample & & 10.5$-$11.0  & &  0.400 & &  0.002 & &  0.150 & &  0.723 & &  0.235 & &  0.673 \\
 \cline{1-17}
                           & &  Voids         & &  9.5$-$10.0  & &  0.400 & &  0.002 & &  0.150 & &  0.723 & &  0.500 & & 0.039 \\
  0.5 $< z <$ 0.7 & &  vs.                    & &  10.0$-$10.5 & &  0.450 & &  $<$ 0.001 & &  0.225 & &  0.231  & &  0.429 & &  0.111\\
                           & & Control Sample & & 10.5$-$11.0  & &  0.425 & &  $<$ 0.001& &  0.250 & &   0.139  & &  0.267 & & 0.589 \\
\enddata
\tablenotetext{}{The output results are based on the $\href{https://docs.scipy.org/doc/scipy/reference/generated/scipy.stats.ks_2samp.html}{\textrm{stats.ks\_2samp (https://docs.scipy.org/doc/scipy/reference/generated/scipy.stats.ks\_2samp.html)}}$ function of python's Scipy package.} 
\tablenotetext{a}{Statistic measures the maximum vertical distance between the empirical cumulative distribution functions of two samples.} 
\tablenotetext{b}{The $p$-value evaluates how well two sample data support the null hypothesis that two distributions are identical.}
\end{deluxetable*}

\begin{figure*}
\includegraphics[scale=1.35]{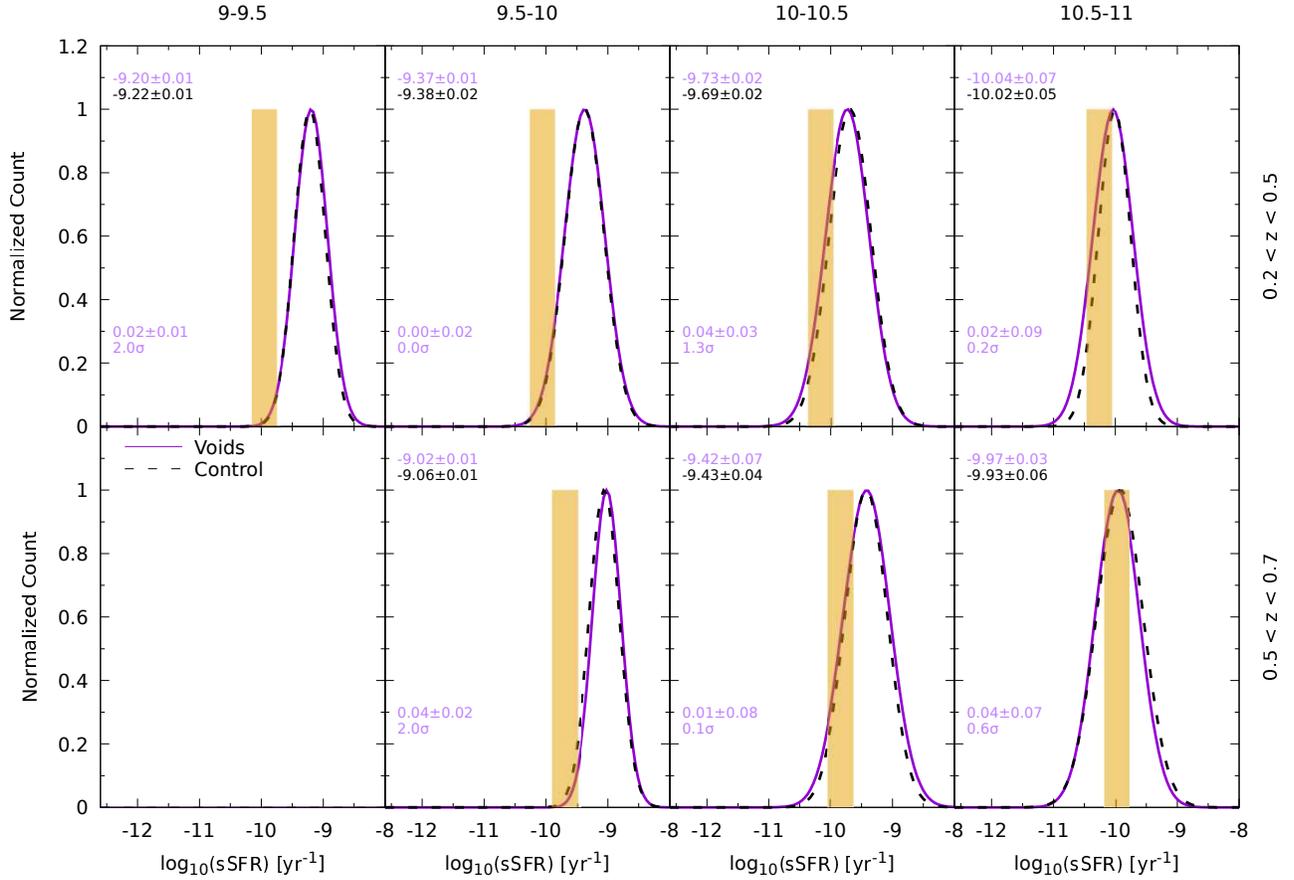}
\caption{The sSFR distributions of the star-forming void and control galaxies. By using a two-Gaussian fit to the sSFR distributions in Figure~\ref{f4}, we decompose the distribution into quiescent and star-forming galaxies for voids (purple) and the control sample (black). The peak values of voids and the control sample are shown at the top left of each subpanel. The differences between the two distributions and their significance are listed at the bottom left. Comparing the distributions of star-forming populations from the voids and the control sample, we find that, in general, the peak difference between the two populations is small ($\leq$ 0.04 dex) with significance $\leq$ 2$\sigma$}, suggesting that the galaxies in voids are very similar to as those in the control sample.
\label{f7}
\end{figure*}

\subsection{sSFR Distributions and Fractions of Quiescent, Star-Forming, and Green Valley Galaxies}\label{sSFRDist}

To further compare the reconstructed void (4) and the control sample (2) in Figure~\ref{f3}, we plot the sSFR distributions as functions of stellar mass and redshift for voids (the solid purple line) and control sample (the dash black line) as shown in Figure~\ref{f5}. We can see that the normalized sSFR profiles evolve with stellar mass and redshift. Besides, the sSFR distributions of void and control galaxies are qualitatively similar, but not the same in detail. Visually, it seems that there is a clear difference for high stellar masses, in good agreement with the finding in \cite{flo21}, but not so much for the lowest mass. We note that the sample of void galaxies consists of roughly one-tenth of the control sample, and the reconstructed void stack is distinct from the whole sample. We also perform a two-side Kolmogorov\textendash Smirnov test (or K\textendash S test) to statistically assess the difference, and list the results in Table~\ref{f3}. In general, we find that the $p$-value is relatively small ($<$ 0.005), implying that two samples are very likely not drawn from the same distribution.   
\begin{figure*}
\includegraphics[scale=1.35]{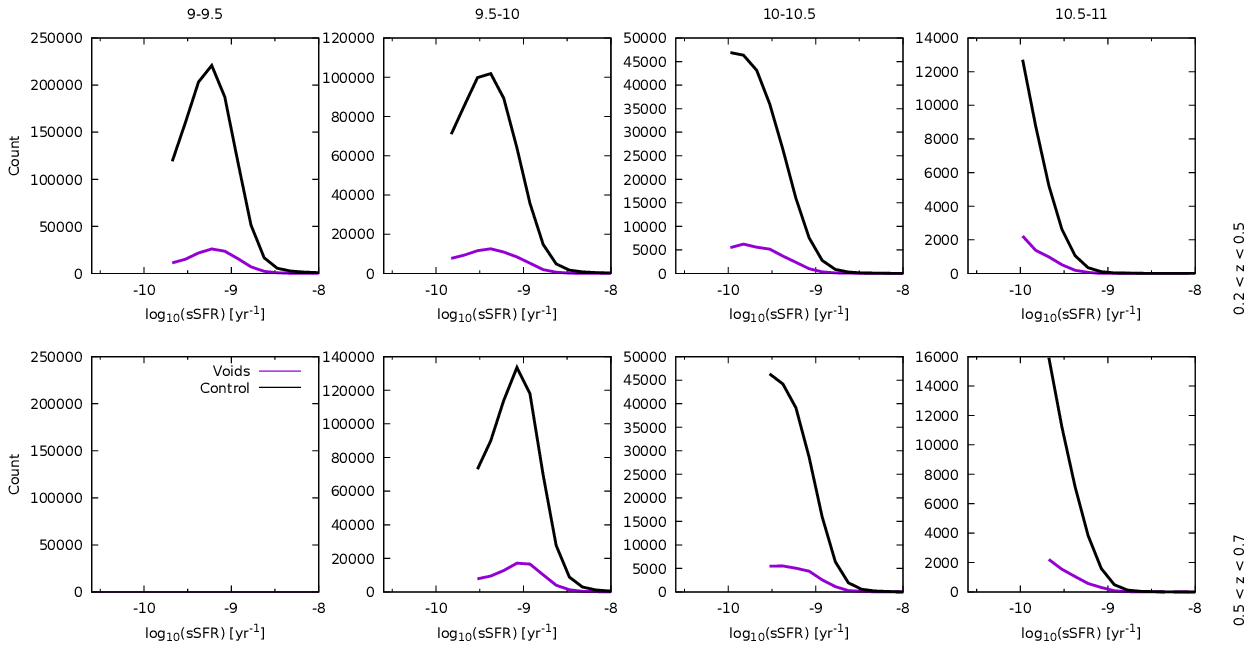}
\caption{The sSFR distribution of the star-forming void and control galaxies based on the classification criterion defined in Section~\ref{denplot}. Additionally, the results of the two-side K\textendash S test are listed in Table~\ref{tab3}}.
\label{f8}
\end{figure*}

Based on the classification method for star-forming, quiescent, and green valley galaxies, discussed in the first paragraph of Section~\ref{denplot}, we measure the fractions of these three populations at two redshift bins, and present the results in Figure~\ref{f6} and list the properties in Table~\ref{tab2}, including the fractions of star-forming ($f_s$), quiescent ($f_q$), and green valley galaxies ($f_g$), as well as the total galaxy number. Besides, we include additional data points for group and cluster galaxies taken from \cite{jian20} for comparison. We see that from the highest to lowest quiescent fraction are galaxies in clusters (orange), groups (green), the control sample (black), and voids (purple). By contrast, for the star-forming and green valley galaxy fractions, the order is reversed, i.e., the fractions of star-forming or green valley galaxies are highest for void galaxies and lowest for cluster galaxies. Additionally, the three fractions exhibit mild redshift evolution in Figure~\ref{f6}.

Moreover, we find that the difference is not only exhibited between the quiescent fractions of void and control galaxies, but is also displayed between the star-forming fractions and green valley galaxy fractions. However, the difference in $f_s$, $f_g$, or $f_q$ is not large, $<$ $\sim$0.1 and its significance is $<$ 2.0$\sigma$. A larger sample would be needed to provide more concrete conclusions. In \cite{flo21}, they found that void galaxies are bluer, more gas-rich, and have more star-forming galaxies than nonvoid galaxies under mass control. Using mock catalogs with and without the assembly bias, they further confirm that their observational results are in good agreement with the mock catalog with an assembly bias, and interpret their finding as a consequence of the assembly bias.

Similar to the finding in \cite{flo21}, we also find more star-forming galaxies (or fewer quiescent galaxies) in voids than in the control sample, although the confidence level of our finding is not statistically strong. The difference in $f_g$ between voids and the control sample is a distinct signature of an environmental dependence. Nevertheless, as we discussed in \cite{jian20}, due to the dominance of quiescent galaxies, the difference in $f_g$ is not able to correctly reflect the frequency of quenching. Instead of $f_g$, we use the effective green valley fraction discussed in the next section to compare the quenching fractions in different environments.

\begin{figure*}
\center
\includegraphics[scale=1.3]{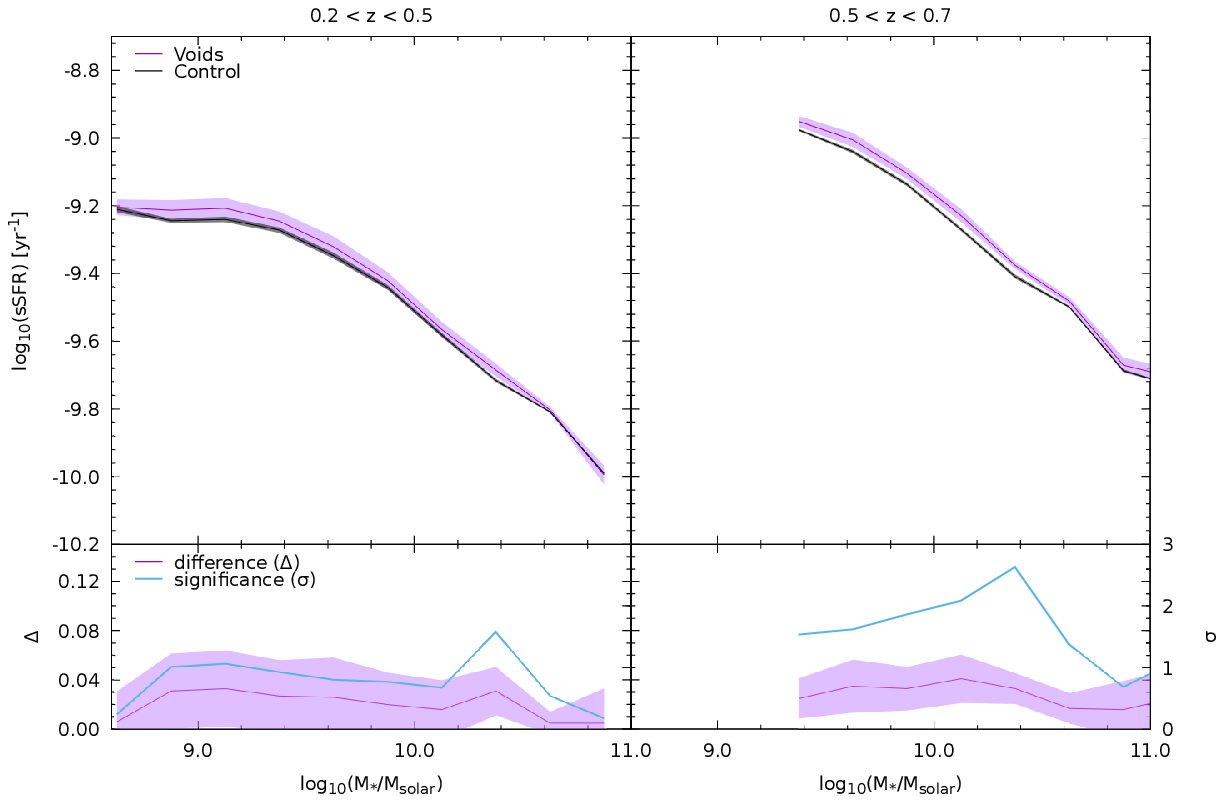}
\caption{The median sSFRs of the star-forming void galaxies for voids (purple) and the control sample (black) are plotted in the top subpanels. Their difference and significance are shown in the bottom subpanels. It appears that the median sSFR of void galaxies deviates slightly from that of control galaxies, with a difference $<$ 0.04 dex and significance $<$ 2.5$\sigma$ for less massive galaxies, in agreement with the finding in Section~\ref{sfssfr}.}
\label{f9}
\end{figure*}

\subsection{sSFR Distributions of Star-forming Galaxies}\label{sfssfr}
As discussed in \cite{jian18,jian20}, the slow environmental quenching effect gradually ceases the star formation of galaxies over longer timescales ($>$1 Gyr). Therefore, we expect that the slow environmental quenching will leave an imprint on star-forming galaxies such that the star-forming main sequence will slightly shift toward low sSFR. Since the void galaxies evolve secularly and are in principle free from the environmental effect, examining the shapes and peak sSFR values of the star-forming main sequences in different environments relative to those of the void galaxies may provide some hints about the environmental quenching effect. To extract the information for the sSFR distribution of star-forming galaxies, we utilize a two-Gaussian profile to decompose the star-forming and quiescent populations, and find the best fit for the whole sSFR profile. In this approach, we do not consider the green valley galaxies as a separate population, but, instead, they represent the tails of star-forming and quiescent populations. The star-forming sSFR distributions as functions of stellar mass and redshift for voids (purple) and the control sample (black), normalized by their peak values, are then plotted in Figure~\ref{f7}. In each subpanel, the peak values of the sSFR profiles for voids and the control sample are shown on the top left. The peak differences between the void and control star-forming galaxies, and their significance, defined in \cite{jian20} as the ratio of the difference to the difference error, are listed at the bottom left.

\begin{figure*}
\center
\includegraphics[scale=1.3]{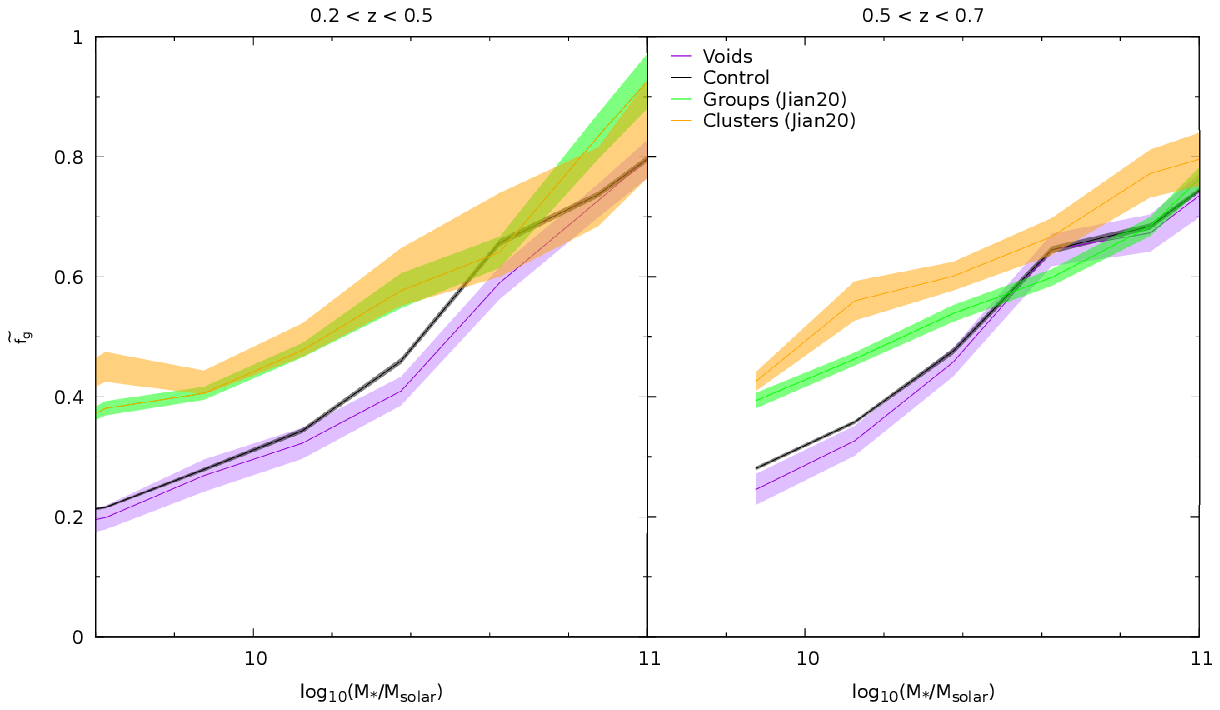}
\caption{The effective green valley galaxy fraction ($\widetilde{f_{g}}$), defined as the fraction of star-forming galaxies number to the nonquiescent galaxies number. We also present two additional data lines for $\widetilde{f_{g}}$ of group (green) and cluster galaxies (orange), adopted from \cite{jian20}. It is apparent that the $\widetilde{f_{g}}$ of the control sample roughly overlaps with that of void galaxies. Generally, the difference is $\leq$ 0.03, and the significance is $\leq$ 1.5$\sigma$, indicating that in terms of $\widetilde{f_{g}}$, the void and control environments are similar, where galaxies suffer similar quenching processes.}
\label{f10}
\end{figure*}

At the low-mass bin, we find that the peak of the void star-forming galaxies is slightly larger than that of the control ones, showing a difference of 0.04 dex in voids with respective to the control sample. The majority of the significance of the difference for star-forming galaxies is less than 1$\sigma$. Additionally, we perform the two-side K\textendash S test for the best-fit distributions of star-forming galaxies in voids and the control sample in Figure~\ref{f7}, and list the results in Table~\ref{tab3}. We find that the $p$-values from all of the mass and redshift bins range from 0.13 to 0.7, indicating that the two distributions are very similar and unlikely to be different.

Alternatively, we select star-forming galaxies based on the classification criterion for the three populations in Section~\ref{denplot} from Figure~\ref{f5} for void and control galaxies, i.e., the data points above the green valley zone, and plot them in Figure~\ref{f8} without the normalization. We again run the K\textendash S test to evaluate the significance of the difference between the two distributions in Figure~\ref{f8}, and present the results in Table~\ref{tab3}. We can see that only a minority of the mass bins show a small $p$-value, and thus the significance of the difference appears to be weak. Based on the K\textendash S tests, we do not obtain strong evidence of the apparent difference between the two distributions, and thus conclude that there is no significant difference between the star-forming galaxies of voids and the control sample, consistent with the findings from \cite{ric14} and \cite{bru20}. 


\subsection{Median sSFR of Star-forming Galaxies}

Another way to quantify the difference between star-forming void and control galaxies is to compare their median sSFRs as a function of stellar mass in these two samples. In Figure~\ref{f9}, the top two subpanels show the median sSFRs of star-forming galaxies for voids (purple) and the control sample (black) at low (left) and high (right) redshift, respectively. The difference of the median sSFR (purple) and its significance (blue) are plotted in the bottom two panels. We can see that, in general, the median sSFRs of the star-forming void and control galaxies seem to depart slightly, but the difference is less than 0.04 dex at a significance level between 1 and 2.5 $\sigma$. This conclusion is consistent with our previous finding in Figure~\ref{f7}.

\subsection{Effective Green Valley Galaxy Fraction}
As discussed in \cite{jian20}, to remove the effect due to the dominance of quiescent galaxies, we define the effective green valley galaxy fraction $\widetilde{f_{g}}$ as the number of green valley galaxies over the number of nonquiescent ones. In Figure~\ref{f10}, $\widetilde{f_{g}}$ is plotted as a function of stellar mass for voids (purple) and the control sample (black) at two redshift bins. We additionally add two data lines representing $\widetilde{f_{g}}$ of groups (green) and clusters (orange), respectively, from \cite{jian20}, for comparison. 

We find that $\widetilde{f_{g}}$ increases with stellar mass no matter what the environment. High-mass galaxies have a higher fractions of green valley galaxies than low-mass ones. In addition, unlike the apparent difference between the group and control galaxies ($\sim$ 0.13) or between the cluster and control ones ($\leq$ 0.2), we can see that $\widetilde{f_{g}}$ of void galaxies roughly coincides with that of control galaxies, especially at high redshift. In general, the difference between $\widetilde{f_{g}}$ in voids and the control sample is approximately $\leq$ 0.03 and the significance is $\leq$ 1.5 $\sigma$, suggesting that there is roughly no deviation between the environments of voids and the control sample in terms of $\widetilde{f_{g}}$. Therefore, we consider galaxies in voids or the control sample to suffer similar physical processes and quenching frequencies, and conclude that the cause of having more star-forming galaxies in voids is not due to different physical processes or quenching frequencies compared to the control sample.

The effect of halos that were formed earlier being more clustered than those with the same mass that were assembled latter is often referred as to ``halo assembly bias''. Similarly, galaxy assembly bias states the correlation between galaxy properties and galaxy formation and assembly history. In order to test the effect of the assembly bias on the properties of void galaxies, \cite{flo21} made use of the conditional abundance matching method to match galaxy properties in the Environmental COntext catalog \citep{mof15,eck16} to (sub)halos in the Vishnu cosmological $N$-body simulation \citep{joh19} with built-in galaxy assembly bias. They concluded that galaxy assembly bias is required to account for their results. In this work, we find no sign of quenching related to the environment between void and control galaxies, but observe more star-forming galaxies in voids than in the control sample, broadly consistent with the finding from \cite{flo21}. Our result thus also favors the assembly bias being the most likely source responsible for more star-forming galaxies in voids than in the control sample. 




\section{Summary}
We utilize the HSC Wide Survey to study galaxy properties, such as the median sSFR of the star-forming main-sequence galaxies and the galaxy fractions for different populations in the BOSS-identified voids. We stack galaxies inside the voids within a redshift slice by subtracting a control stack, build by selecting all of the galaxies in the HSC survey field in the same redshift range, from the void stack to recover the intrinsic void galaxy properties statistically. Galaxies are classified into star-forming, green valley, and quiescent populations on the SFR\textendash $M_{*}$. Comparisons of galaxy properties are then made between the void and control environments in two redshift bins ($0.2<z<0.5$ and $0.5<z<0.7$). The main results are summarized as follows:

\begin{enumerate}
 
 \item The revised voids by including faint galaxies in the same area of the voids defined by using bright galaxies still reveal underdense density relative to the density of the control sample.   
 \item From the sSFR distributions of galaxies in the void and control environments, we observe that the two profiles are similar but not the same. Void galaxies seem to have more star-forming galaxies than the control galaxies. Besides, from the two-side K\textendash S tests, the $p$-values also reveal that the sSFR distributions of void and control galaxies are statistically different.
 \item By comparing the sSFR distribution of star-forming galaxies and the median sSFR of star-forming galaxies, we show that star-forming void galaxies appear to be similar to star-forming control galaxies.
 \item When we split the galaxies into three populations, we observe a higher fraction of star-forming galaxies in voids than in the control sample at fixed mass and redshift at a 2$\sigma$ level. A larger sample will be required to confirm this trend. In addition, the intrinsic fractions of the green valley galaxies in voids are slightly higher than those in the control galaxies. The assembly bias and environmentally associated processes are two possible causes of the observed difference in the fractions of quiescent and green valley galaxies.
 
 \item To differentiate between the assembly bias and environmental effects, which are responsible for the greater fraction of star-forming galaxies seen in the voids, we calculate the effective green valley galaxy fraction, defined as the number of green valley galaxies divided by the number of nonquiescent galaxies. We find that the effective green valley galaxy fraction in voids agrees well with that in the control sample. The difference is small and may be neglected, implying galaxies likely suffer similar physical processes and quenching frequencies between the voids and the control sample. That is, the observed difference in the fractions of star-forming galaxies and green valley galaxies is unlikely due to processes associated with the environments. In other words, under the mass and redshift control, we find no environmental effects, but more star-forming galaxies in voids. Considering all the results together, our result thus favors the scenario of galaxy assembly bias.     

\end{enumerate}

Cosmic voids are enormous structures in the universe and therefore a large survey area is required to properly identify enough cosmic voids for statistical analysis. While the overlapping area of the HSC survey with the BOSS void sample used in this work is not sufficiently large, its profound depth improves upon previous works by exploring the characteristics of faint void galaxies. Our results reveal that galaxies residing in underdense environments are dissimilar from those living in overall surroundings for masses down to $\sim$ 10$^{9.6}$ solar mass at $z$ $\sim$ 0.2. For the future perspective, a deep and wide photometric survey covering the entire SDSS DR12 BOSS field will help us to probe faint void galaxies with sufficient statistics and better constrain the difference and significance between void and overall galaxies.




\emph{Acknowledgments}-
We thank the anonymous referee for his/her helpful comments and suggestions. L.L. is thankful for the support from Academia Sinica under the Career Development Award CDA-107-M03 and for the Ministry of Science and Technology of Taiwan under the grant MOST 108-2628-M-001-001-MY3. K.U. acknowledges support from the Ministry of Science and Technology of Taiwan (grants MOST 106-2628-M-001-003-MY3 and MOST 109-2112-M-001-018-MY3) and from the Academia Sinica Investigator Award (grant AS-IA-107-M01).

The Hyper Suprime-Cam (HSC) collaboration includes the astronomical communities of Japan and Taiwan, and Princeton University.  The HSC instrumentation and software were developed by the National Astronomical Observatory of Japan (NAOJ), the Kavli Institute for the Physics and Mathematics of the Universe (Kavli IPMU), the University of Tokyo, the High Energy Accelerator Research Organization (KEK), the Academia Sinica Institute for Astronomy and Astrophysics in Taiwan (ASIAA), and Princeton University.  Funding was contributed by the FIRST program from the Japanese Cabinet Office, the Ministry of Education, Culture, Sports, Science and Technology (MEXT), the Japan Society for the Promotion of Science (JSPS), Japan Science and Technology Agency  (JST), the Toray Science  Foundation, NAOJ, Kavli IPMU, KEK, ASIAA, and Princeton University.

This paper makes use of software developed for the Large Synoptic Survey Telescope. We thank the LSST Project for making their code available as free software at  http://dm.lsst.org

This paper is based (in part) on data collected at the Subaru Telescope and retrieved from the HSC data archive system, which is operated by Subaru Telescope and Astronomy Data Center (ADC) at NAOJ. Data analysis was in part carried out with the cooperation of Center for Computational Astrophysics (CfCA), NAOJ.

The Pan-STARRS1 Surveys (PS1) and the PS1 public science archive have been made possible through contributions by the Institute for Astronomy, the University of Hawaii, the Pan-STARRS Project Office, the Max Planck Society and its participating institutes, the Max Planck Institute for Astronomy, Heidelberg, and the Max Planck Institute for Extraterrestrial Physics, Garching, The Johns Hopkins University, Durham University, the University of Edinburgh, the Queen’s University Belfast, the Harvard-Smithsonian Center for Astrophysics, the Las Cumbres Observatory Global Telescope Network Incorporated, the National Central University of Taiwan, the Space Telescope Science Institute, the National Aeronautics and Space Administration under grant No. NNX08AR22G issued through the Planetary Science Division of the NASA Science Mission Directorate, the National Science Foundation grant No. AST-1238877, the University of Maryland, Eotvos Lorand University (ELTE), the Los Alamos National Laboratory, and the Gordon and Betty Moore Foundation.

\end{document}